# New high-pressure phase and equation of state of $Ce_2Zr_2O_8$

D. Errandonea[1,*], R. S. Kumar[2], S. N. Achary[3], O. Gomis[4], F. J. Manjón[5], R. Shukla[3] and A. K. Tyagi[3]

[1]*MALTA Consolider Team, Departamento de Física Aplicada-ICMUV, Universitad de Valencia, Edificio de Investigación, c/Dr. Moliner 50, Burjassot, 46100 Valencia, Spain*

[2]*High Pressure Science and Engineering Center, Department of Physics and Astronomy, University of Nevada Las Vegas,4505 Maryland Parkway, Las Vegas, Nevada 89154-4002, USA*

[3]*Chemistry Division, Bhabha Atomic Research Centre, Trombay, Mumbai 400085, India*

[4]*Centro de Tecnologías Físicas: Acústica, Materiales y Astrofísica, MALTA Consolider Team, Universitat Politècnica de València, 46022 València, Spain*

[5]*Instituto de Diseño para la Fabricación y Producción Automatizada, MALTA Consolider Team, Universitat Politècnica de Valencia, 46022 València, Spain*

**Abstract:** In this paper we report a new high-pressure rhombohedral phase of $Ce_2Zr_2O_8$ observed from high-pressure angle-dispersive x-ray diffraction and Raman spectroscopy studies up to nearly 12 GPa. The ambient-pressure cubic phase of $Ce_2Zr_2O_8$ transforms to a rhombohedral structure beyond 5 GPa with a feeble distortion in the lattice. Pressure evolution of unit-cell volume showed a change in compressibility above 5 GPa. The unit-cell parameters of the high-pressure rhombohedral phase at 12.1 GPa are $a_h$ = 14.6791(3) Å, $c_h$ = 17.9421(5) Å, V = 3348.1(1) Å$^3$. The structure relation between the parent cubic ($P2_13$) and rhombohedral ($P3_2$) phases were obtained by group-subgroup relations. All the Raman modes of the cubic phase showed linear evolution with pressure with the hardest one at 197 cm$^{-1}$. Some Raman modes of the high-pressure phase have a non-linear evolution with pressure and softening of one low-frequency mode with pressure is found. The compressibility, equation of state, and pressure coefficients of Raman modes of $Ce_2Zr_2O_8$ are also reported.

[*] Author to whom correspondence should be addressed. Electronic mail: daniel.errandonea@uv.es.

## 1. Introduction

$A_2B_2O_7$ pyrochlores are subject of great interest due to their oxygen mobility[1] and catalytic properties,[2] as well as for their use as stable and robust host lattice for nuclear-waste immobilization[3] and inert-matrix fuel for nuclear technology.[4] The preferential coordination of A and B cations in pyrochlores restricts their ionic-radii ratio between 1.46 to 1.8.[5] Thus, pyrochlores are formed only with cations of appropriate ionic-radii and charge combinations. Smaller radii ratios favor an anion-deficient fluorite lattice, but this limit can be extended under high-pressure (HP).[6] Among pyrochlores, rare-earth zirconates are being extensively investigated under HP due to their interest for nuclear-waste storage. Information obtained from these studies is also valuable for their use as host lattice for transmutation products of reactors and safe deposition of plutonium.[3,7,8] The high radiation and chemical stability as well as low neutron absorption coefficients of Zr make zirconate pyrochlores extremely useful for these purposes, but recent studies found pressure-driven unstabilities.[9] Structural transitions were reported for $Sm_2Zr_2O_7$, $Gd_2Zr_2O_7$, $Er_2Zr_2O_7$, and $Ho_2Zr_2O_7$ at 15 - 30 GPa.[10-12] Studies on related $Y_2Zr_2O_7$ indicated the formation of a defect-cotunnite structure at 18 GPa.[13] HP x-ray diffraction (XRD) and Raman studies of $La_2Hf_2O_7$ revealed no phase transition instead showed a decomposition to constituent oxides.[14] Similar studies of $Gd_2Zr_2O_7$ detected a monoclinic $La_2Ti_2O_7$-type phase at significantly lower pressure, ~ 3 GPa.[15] HP studies on several analogues found a pressure-induced amorphization.[14,16-19]

Though a number of studies exist on the HP behavior of rare-earth zirconate pyrochlores, few reports are available on $Ce_2Zr_2O_7$. Since cerium oxide ($CeO_2$) is used as a surrogate material to simulate thermo-physical properties of radioactive $PuO_2$, the studies on cerium containing oxides are extremely important for nuclear technology. Recently, Surble *et al.*[20] have investigated $Ce_2Zr_2O_7$ reporting a structural transition at 20 GPa.[20] $Ce_2Zr_2O_7$ is a metastable material at ambient conditions and gradually oxidizes to $Ce_2Zr_2O_{7+x}$ in air.[21-23] Studies on oxidized samples of $Ce_2Zr_2O_7$ revealed several new phases.[24-29] In particular, $Ce_2Zr_2O_8$ exists as cubic fluorite structure[29,30] or metastable tetragonal structures.[31,32]

It is also well known that the fluorite structure transforms to different crystalline modifications with pressure.[33] HP studies on $CeO_2$ suggest the fluorite to orthorhombic $\alpha$-$PbCl_2$-type structure transition. The high-pressure phase relaxes to a hexagonal structure on decompression.[34,35] HP studies on $ZrO_2$ and $HfO_2$ indicate the formation of orthorhombic or monoclinic phases.[35,36,37] HP studies on columbite-type $ZrTiO_4$ ($HfTiO_4$) show the formation of baddeleyite-related HP phases above 9 (2.7) GPa.[38] The diverse possible structural transitions of fluorite, zirconia, and pyrochlore create more interest on the HP behavior of $Ce_2Zr_2O_8$, which represents a close combination of fluorite and pyrochlore. In order to further understand the HP behavior of these structures, we have carried out high-pressure XRD and Raman studies on $Ce_2Zr_2O_8$. Its compressibility and equation of state (EOS) are obtained. In addition we report a new non quenchable HP phase of $Ce_2Zr_2O_8$.

## 2. Experimental details

The $Ce_2Zr_2O_8$ sample used in the experiments was prepared by gel-combustion method from the corresponding metal nitrates [$Ce(NO_3)_3 \cdot 6H_2O$, 99.999 %, Alfa Aesar; $ZrO(NO_3)_2 \cdot xH_2O$, 99.99 %, Aldrich] using glycine as a fuel. The metal contents of the reactants were standardized by thermogravimetric analysis of the corresponding nitrates. Appropriate amounts of nitrates were mixed together in demineralised water and a stoichiometric amount of glycine was added to the solution. Highly viscous gel was obtained by slow dehydration of the solution at ~ 80 ºC on a hot plate. On further raising the temperature the gel undergoes an auto-ignition process with evolution of a large volume of gases and finally converted to voluminous powder. The light-yellow powder obtained was calcined at 900 ºC for about 4 h to remove any residual un-burn carbonaceous material. The calcined powder was pressed into pellets and heated at 1400 ºC for 48 h in flowing atmosphere of Argon-$H_2$ (8 % v/v hydrogen-in argon) gas mixture, with intermittent grinding. From powder XRD data, the black powder obtained after this treatment was confirmed as pyrochlore-type $Ce_2Zr_2O_7$. This powder was slowly heated to 1000 ºC for 5 h in air to completely oxidize all $Ce^{3+}$ to $Ce^{4+}$. The bright-yellow oxidized product was characterized by XRD and neutron diffraction (ND) confirming the $Ce_2Zr_2O_8$ composition. The XRD pattern at ambient temperature was recorded on Panalytical X-Pert pro powder X-ray diffractometer using Cu $K_\alpha$ radiation. The ND data were recorded at Dhruva Research reactor, Trombay, Mumbai using neutrons of wavelength 1.249 Å.

In order to perform HP studies a finely ground powder sample was used. Synchrotron powder x-ray diffraction experiments were performed up to 12.1 GPa. Pre-pressed pellets obtained from the powder were loaded in a 100-μm hole of a rhenium gasket pre-indented to 30 μm in a diamond-anvil cell (DAC) with diamond-culet sizes of 300 μm. Ruby grains were loaded with the sample for pressure determination[39] and neon (Ne) was used as the pressure-transmitting medium.[40,41] At pressures higher than 4 GPa (solidification of Ne) the EOS of Ne was used to double check the pressure.[42] Differences in the measured pressure with different scales were always smaller than 0.2 GPa. Angle-dispersive x-ray diffraction (ADXRD) experiments were carried out at Sector 16-IDB of the HPCAT, at the Advanced Photon Source (APS), with an incident wavelength of 0.4246 Å. The monochromatic x-ray beam was focused down to 10×10 $\mu m^2$ using Kickpatrick-Baez mirrors. The images were collected using a MAR345 image plate located 350 mm away from the sample and then integrated and corrected for distortions using FIT2D.[43] The structural analyses and refinements were performed using the POWDERCELL[44] and Fullprof-2000[45] program packages.

For high-pressure Raman spectroscopic studies a pre-pressed powder sample along with 2-μm diameter ruby balls was loaded in a pre-indented steel gasket with a 200-μm diameter hole inside a DAC. A 16:3:1 methanol-ethanol-water mixture was used as pressure-transmitting medium.[40,46] The pressure was determined by monitoring the shift in ruby fluorescence lines.[39] HP Raman measurements were performed in the backscattering geometry using a 632.8 nm HeNe laser and a Horiba Jobin Yvon LabRAM HR UV microspectrometer in combination with a thermoelectric-cooled multichannel CCD detector with spectral resolution below 2 $cm^{-1}$.

## 3. Results and discussion

The phase purity of $Ce_2Zr_2O_8$ was confirmed from Rietveld refinement of ND data. For its structure we considered distinct positions for two Ce (*4a* and *12b* sites), two Zr (*4a* and *12b* sites), and eight oxygen atoms (four in *4a* and four in *12b* sites) in a cubic ($P2_13$) lattice. The refined unit-cell parameters, $a$ = 10.5444(2) Å and V = 1172.37(4) Å$^3$, agree with the literature.[29] The structure of $Ce_2Zr_2O_8$ is represented in **Fig. 1**; its structural parameters are given in **Table I**. As mentioned above, the structure is closely related to parent $Ce_2Zr_2O_7$ pyrochlore except that the empty anion positions are filled with additional oxygen atoms. The cation and anion sites of $Ce_2Zr_2O_7$ are split in $Ce_2Zr_2O_8$ by lowering the symmetry from $Fd\bar{3}m$ to $P2_13$. Cations form distorted $CeO_8$ and $ZrO_8$ polyhedra connected by sharing their edges as in fluorite.

XRD patterns recorded at different pressures are shown in **Fig 2**. All the reflections in XRD patterns recorded from ambient pressure to 5.1 GPa could be indexed on a primitive cubic lattice similar to the one observed at ambient pressure outside the DAC. The full patterns were further Rietveld refined to get the structural parameters. Typical observed and calculated diffraction patterns of $Ce_2Zr_2O_8$ at a representative pressure are depicted in **Fig 3**. The refined unit-cell parameters for cubic ($P2_13$) $Ce_2Zr_2O_8$ at different pressures are listed in **Table II**. On further compression the XRD data appear quite similar to the $Ce_2Zr_2O_8$ lattice except from weak peaks that emerge at 8.34º, 11.83º, 12.87º, and 22.94º. These peaks are marked in **Fig 2**. They become prominent at 12.1 GPa. The Le Bail refinement of the XRD pattern using the space group $P2_13$ can account for all the observed intense reflections. However, the weak new reflections remain un-indexed in the calculated pattern. The calculated cubic unit-cell parameters of these high-pressure XRD data were included in **Table II**. The parameters obtained by such refinement indicate that the structure is grossly cubic and the weak reflections appear due to a feeble structural distortion in the lattice. The variation of unit-cell volume with pressure is shown in **Fig 4**. The linear and volume compressibilities of $Ce_2Zr_2O_8$ within different pressure regions are given in **Table II**. The average axial compressibility ($1.26 \times 10^{-3}$ $GPa^{-1}$) is comparable to that reported for $Ce_2Zr_2O_7$ ($1.3 \times 10^{-3}$ $GPa^{-1}$)[19]. It is also observed that the compressibilities calculated from the unit-cell parameters up to 5.1 GPa are higher than those calculated from the data beyond 5.1 GPa. This is also evident from the deviation in the variation of the unit-cell volume (**Fig. 4**). The pressure-volume data obtained from both pressure ranges were fitted with the third order Birch-Murnaghan EOS.[47]

$$P = \frac{3B_0}{2}\left[\left(\frac{V_0}{V}\right)^{7/3} - \left(\frac{V_0}{V}\right)^{5/3}\right] \times \left(1 + \frac{3}{4}(B_0' - 4)\left(\left(\frac{V_0}{V}\right)^{2/3} - 1\right)\right)$$

The obtained EOS parameters are $V_0$ = 1172(2) Å$^3$, $B_0$ = 214(5) GPa, and $B_0$' = 8(1), where $V_0$ is the unit-cell volume at ambient pressure, $B_0$ the bulk modulus at ambient pressure, and $B_0$' its pressure derivative. The bulk modulus is slightly larger than that of other pyrochlore-type zirconates summarized in **Table III**,[14,15,20,48] but significantly smaller compared to fluorite- and cotunnite-type lattices[34] (see **Table III**). Direct comparison with $Ce_2Zr_2O_7$ cannot be done because an unphysical value for $B_0$' is reported for its EOS.[20] The bulk-modulus enhancement in comparison with $A_2B_2O_7$ pyrochlores can be attributed to the complete filling with oxygen of the vacant *8a* sites of the pyrochlore ($Fd\bar{3}m$) lattice. The decrease in comparison with fluorites can be caused by the relatively lower packed lattice of $Ce_2Zr_2O_8$ compared to them. The vacancy filling can be thought as an internal pressure, which makes $Ce_2Zr_2O_8$ to behave as a compressed version of $Ce_2Zr_2O_7$ reducing compressibility and the transition pressure. Another interesting fact to note is that a large $B_0$' has been observed for $Ce_2Zr_2O_8$ and also for most pyrochlores studied so far. In the present case, this fact is a consequence of the compressibility change observed at the phase transition. This suggests that probably similar transitions (and compressibility changes) could take place in pyrochlores that have not been previously detected given the subtle characteristics of the cubic-rhombohedral transition. A possible cause of the missing transition could be the use of less hydrostatic pressure media than Ne.

Further analysis of the weak reflections in high-pressure XRD data confirms a phase transition in $Ce_2Zr_2O_8$. In most of the zirconate pyrochlores structural transitions at high pressure (~ 20 GPa) have been reported. The phase-transition pressure (~ 7 GPa) observed in the present case is significantly lower than that expected for pyrochlore-type phases. The HP phases observed in pyrochlores are often explained by distorted defect-fluorite-type or cotunnite-type structures or monoclinically distorted $ZrO_2$-type or $La_2Ti_2O_7$-type structures[9,15,20]. The cation diffusion to form these structures can only be expected and observed at higher pressures than 7 GPa. The transition at low pressure might be related to a distortion in the lattice without any significant redistribution of cations or anions. Thus, the HP phase transition in $Ce_2Zr_2O_8$ can be related to a distortion in the lattice instead of any significant change in the frame of the lattice.

Further evidence of phase transition could be obtained from HP Raman studies. The Raman spectra of $Ce_2Zr_2O_8$ recorded at different pressures are shown in **Fig. 5**. The Raman spectrum at ambient pressure is quite similar to that earlier reported by Matsuo *et al.*[49] According to group-theory analysis $Ce_2Zr_2O_8$ has 144 Raman active modes ($\Gamma$ = 24 A + 24 $E^1$ +24 $E^2$ + 72 T). Previously only seven modes have been reported for cubic $Ce_2Zr_2O_8$.[48] Here, twenty-one modes could be assigned clearly at ambient-pressure (marked in **Fig. 5).** The pressure evolution of Raman spectra indicates the appearance of new bands above 5.5 GPa (see the spectra measured at 5 and 5.5 GPa in **Fig. 5**) in addition to the bands observed at ambient pressure. The intensities of these new bands gradually increase upon compression. From 6.7 to 12.6 GPa the Raman spectra can be assigned to the HP phase. The reversibility of the transition is concluded from the similarity of the pressure-released Raman spectrum and the original ambient-pressure spectrum. This further supports the observed phase transition in XRD studies.

Systematic HP studies on pyrochlore-type $A_2B_2O_7$ lattices indicate commonly the pyrochlore ($Fd\bar{3}m$) to defect-fluorite-type ($Fm\bar{3}m$) or defect-cotunnite-type (Pnma) transitions. They are observed with a concomitant cation disordering and finally as a geometrically frustrated amorphous phase. A transition induced by a structural distortion at low pressure, viz 3 GPa, has been only reported in $Gd_2Zr_2O_7$.[15] Systematic studies on the fluorite-type lattices also indicate the structural instability of fluorite structures and transformation from cubic fluorite-type to cotunnite-type lattices at higher pressures. HP studies on cubic $CeO_2$ indicate that a structural transition to orthorhombic $PbCl_2$-type lattice occur at 31 GPa[50,51]. The same transition is reported in nanocrystalline $CeO_2$ at 22.3 GPa.[34,35] A large volume collapse with an increase in coordination number in the metal ions is observed at this transition. The pressure-induced orthorhombic columbite-type $ZrTiO_4$ and $HfTiO_4$ to monoclinic baddeleyite-type structures occur at relatively lower pressure.[37] The monoclinic baddeleyite-type $ZrO_2$ transforms to orthorhombic cotunnite-type $ZrO_2$ at high pressure.[36] Leger *et al.* have reported the sequence of structural transitions as baddeleyite - orthorhombic-I (Pbca) - orthorhombic-II - orthorhombic-III at room temperature around 10, 25, and 42 GPa, respectively [35]. Thus, it is not uncommon to expect a pressure-induced structural distortion in $Ce_2Zr_2O_8$. The change in compressibilties and appearance of new peaks in XRD and Raman data clearly suggest a phase transition in $Ce_2Zr_2O_8$.

Refinement of XRD data obtained at 12.1 GPa shows a clear difference from that of the parent cubic $Ce_2Zr_2O_8$ (**Fig. 6**). The weak reflection observed at 11.83° can be assigned to (111) reflection of solid Ne.[42] The analysis of other extra reflections which are weak could not be accounted to any of $CeO_2$ or $ZrO_2$ ambient or high-pressure phases (or their super structures). The transformation of zirconate pyrochlores to defect fluorite-type and defect cotunnite-type phases is generally observed from the appearance of a satellite reflection adjacent to the intense (222) reflections.[9] However the appearance of other extra reflections and the prominent peak at 22.9° could not be accounted by such phases. Thus we understand the weak reflections originate from a feeble structural distortion of cubic $Ce_2Zr_2O_8$.

In order to study the new phase formed at HP we have tried to index all the reflections including the weak ones observed at 12.1 GPa. They could be indexed on a hexagonal lattice with lattice parameters as: $a_h$ = 14.674(9) Å and $c_h$ = 17.83(3) Å. The observed unit-cell parameters are quite similar to rhombohedral lattice reported by Thomson *et al.* ($R\bar{3}m$, $a_r$ = 10.5439 Å and α = 90.05; $a_h$ = 14.9178 Å and $c_h$ = 18.2466 Å) for oxygen-rich $Ce_2Zr_2O_{7.97}$.[25] This rhombohedral phase has also been prepared by a slow oxidation of pyrochlore-type $Ce_2Zr_2O_7$ similar to the preparation of $Ce_2Zr_2O_8$ of this study. The cations usually do not diffuse under such mild oxidation conditions except the anions are filled in the empty fluorite equivalent sites. It can be mentioned here that the observed unit-cell parameters can be related to the parent cubic unit-cell parameters by matrix transformation as:

$$\begin{pmatrix} a_h \\ b_h \\ c_h \end{pmatrix} = \begin{pmatrix} 1 & -1 & 0 \\ 0 & 1 & -1 \\ 1 & 1 & 1 \end{pmatrix} \times \begin{pmatrix} a_c \\ b_c \\ c_c \end{pmatrix}$$

Where $a_h = b_h$ and $c_h$ are hexagonal parameters of the transformed unit cell and $a_c = b_c = c_c$ are parent cubic unit-cell parameters.

Using the present observed unit-cell parameters with the space group and position coordinates reported by Thomson *et al.* for $Ce_2Zr_2O_{7.97}$ [25] we tried to model the XRD pattern observed at 12.1 GPa. However, the weak reflections observed at 8.34° and 22.9° were excluded from the indexed results, which suggests the absence of rhombohedral (R) centering. Thus the model structure for the observed HP phase was generated by group-subgroup relations from the $P2_13$ lattice of $Ce_2Zr_2O_8$. Excluding the R-centered rhombohedral subgroup of the $P2_13$ lattice, three possible subgroups, namely P3, $P3_1$ and $P3_2$ could be assigned to the HP phase of $Ce_2Zr_2O_8$. The Le Bail refinement of the observed XRD data using $P3_2$ showed a good profile match in all the reflections (**Fig. 7**). The residuals of the refinement are $R_p$ = 4.4%, $R_{wp}$ = 6.7% and $\chi2$ = 0.4. The obtained unit-cell parameters are: $a_h$ = 14.6791(3) Å, $c_h$ = 17.9421(5) Å, and V = 3348.1(1) $Å^3$. There is no volume change between the cubic and rhombohedral structure. The structure of the HP phase is schematically represented in **Fig. 1**.

Considering the close relation between the parent cubic lattice and the transformed rhombohedral ($P3_2$) lattice, it can be suggested that the unit cell contains 24 units of $Ce_2Zr_2O_8$. The structural model for the rhombohedral lattice is proposed from the equivalent position coordinates generated using the transformation matrix. Similar to the cubic lattice, all the metal ions in the rhombohedral lattice retain the cubic coordination with oxide ions. Full structural refinement of the powder XRD data collected from the DAC are usually difficult. In addition the present proposed structure contains large number atoms in asymmetric sites, which makes the analysis even more difficult. Therefore, we have imposed a constrained refinement where all the metal atoms have cubical coordination by bond length as obtained in the derived structure model. The background, profile parameters, and unit-cell parameters obtained from the Le Bail refinement were used for Rietveld refinements. We consider 16 Ce, 16 Zr, and 64 oxygen atoms in the general *3a* positions of space group $P3_2$. Only one overall thermal parameter was used for the refinement. The position coordinates of heavier Ce atoms were refined first and subsequently the Zr and oxygen atoms were added in the refinement. The final Rietveld refinement plot shows small intensity differences in the fundamental reflections, which suggests the grains are oriented in the DAC. This is consistent with the fact that collected patterns for the HP phase tended to be rather spotty. Overall the refinement could provide satisfactory structural parameters. The refined structural parameters are given in **Table IV**.

Before closing the discussion of the crystalline structure of the HP rhombohedral phase we would like to comment on the essential difference between it and the rhombohedral structure obtained at ambient pressure by oxidation.[25] The anion filled rhombohedral pyrochlore phase reported by Thomson *et al.*[25] has a $Ce_2Zr_2O_{7.97}$ composition. Also it has been reported that the sample appear close to yellow orange, which has been attributed to incomplete oxidation of $Ce^{3+}$ as the oxidation is carried out at relatively lower temperature. In the present study the color of the sample is bright yellow, which is the typical color of a fully oxidized sample.[27-29] The structures of the anion filled pyrochlores are closely similar and all of them are related to the parent cubic pyrochlore lattice.

The ambient pressure rhombohedral phase has been explained with distinguishable four Ce, four Zr, and eleven O atoms. Several under-occupied sites for tetrahedral anions and a new under-occupied trigonal site are present in this structure. In addition the ambient rhombohedral structure involves trigonal sites equivalent to the *32e* sites of the cubic pyrochlore lattice, which are displaced anions of the normal *8b* oxygen site. The rhombohedral structure of the high-pressure polymorph is more closely related to the cubic ($P2_13$) phase where no displaced anions are involved. The spitting of *8a* vacant sites and other cation and anion sites of the parent cubic ($Fd\bar{3}m$) $Ce_2Zr_2O_7$ are observed in cubic ($P2_13$) structure. The present reported HP rhombohedral structure have anions sites equivalent to normal anion sites of the $Fd\bar{3}m$ lattice. In both the ambient- and high-pressure rhombohedral structure the cation polyhedra are distorted. In the ambient-pressure structure some of the metal ions show coordination numbers larger than eight with a wider dispersion of bond lengths, viz. Ce-O bonds in the range of 2.03 - 2.70 Å and Zr-O bonds in the range of 2.04 - 2.41 Å. However, in the HP structure all the metal ions retain a cubic coordination with oxide ions, being the eight Ce-O bonds in the 2.25 – 2.48 Å range and the eight Zr-O bonds in the 2.05 – 2.26 Å range; i.e the polyhedral units are comparatively less distorted than in the ambient rhombohedral phase. The HP structure retains the polyhedral framework of the cubic phase, being only gradual distortions of polyhedra induced by the displacive transition.

To close this work, we would like to provide additional information on the pressure evolution of the Raman spectra. From the measured spectra we obtained the pressure evolution of the Raman modes for the low- and high-pressure phases. Results are summarized in **Fig. 8**. All modes show a linear evolution upon compression, being the high-frequency modes the hardest modes and the mode located at 197 $cm^{-1}$. None of the modes has a soft-mode behavior. In **Table V** we report the Raman frequencies (ω) and pressure coefficients (dω/dP) obtained for each mode. The mode Grüneisen parameters are calculated by relation $\gamma = (B_0/\omega)\cdot d\omega/dP$ using $B_0$ = 214 GPa and they are given in **Table V**. For the HP phase we found an increase of the number of Raman modes which is consistent with the cubic-to-rhombohedral symmetry decrease associated to the phase transition. In total we detected thirty-four Raman-active phonons. Most of them have a similar evolution as the phonons of the low-pressure phase. They are also located in the same frequency region. Both facts are consistent with the structural similitude between the low- and high-pressure phases. The frequencies and pressure coefficients are summarized in **Table VI**. It is interesting to note that in the HP phase we found a few modes that follow a non-linear evolution upon compression and at least one mode at low frequencies which soften under compression (see **Fig. 8** and **Table VI**). These facts could be related to a mechanical instability of the HP phase, suggesting the possible occurrence of further transitions at higher pressures.

## 4. Concluding Remarks

A systematic HP study on the cubic $Ce_2Zr_2O_8$, an oxygen-filled pyrochlore analogue, was performed. HP ADXRD studies were carried out in a diamond-anvil cell using synchrotron x-ray and Ne as pressure-transmitting medium. HP Raman studies were carried out in a DAC using methanol-ethanol-water as pressure-transmitting medium. From both experiments we observed a reversible cubic-to-rhombohedral structural transition above 5 GPa. The structural details of the HP rhombohedral phase were obtained by a group-subgroup relation. The structural transition observed in this study is different from usual observations in pyrochlore-type or fluorite-type materials. The HP rhombohedral phase also differs from ambient-pressure rhombohdral oxygen-rich $Ce_2Zr_2O_{7.97}$. We think that this transition, not previously observed in related compounds, is produced by a mechanical stress-induced distortion in the lattice. All the Raman modes detected for the cubic phase at ambient pressure showed a linear evolution with pressure while some of the Raman modes of the HP phase showed a non-linear evolution and even one mode softens with increasing pressure.

**Acknowledgments:** The authors thank the financial support from Spanish MICCIN (Grants MAT2010-21270-C04-01/04 and CSD2007-00045). This work was performed at HPCAT, Advanced Photon Source (APS), Argonne National Laboratory. HPCAT is supported by CIW, CDAC, UNLV and LLNL through funding from DOE-NNSA, DOE-BES and NSF. APS is supported by DOE-BES, under DEAC02-06CH11357. The UNLV HPSEC is supported by the U.S. DOE, National Nuclear Security Administration, under DE-FC52-06NA26274.

**Table I:** Ambient pressure structural parameters as obtained from powder neutron-diffraction data of $Ce_2Zr_2O_8$.

| Atoms | Wyckoff positions | x | y | z | B (Å$^2$) |
|---|---|---|---|---|---|
| $Ce_1$ | *4a* | 0.1355(16) | 0.1355(16) | 0.1355(16) | 0.024 |
| $Ce_2$ | *12b* | 0.1353(16) | 0.3647(16) | 0.3676(16) | 0.021 |
| $Zr_1$ | *4a* | 0.6311(10) | 0.6311(10) | 0.6311(10) | 0.021 |
| $Zr_2$ | *12b* | 0.6232(10) | 0.8748(11) | 0.8670(8) | 0.021 |
| $O_1$ | *4a* | -0.0002(17) | -0.0002(17) | -0.0002(17) | 0.090 |
| $O_2$ | *4a* | 0.2569(10) | 0.2569(10) | 0.2569(10) | 0.090 |
| $O_3$ | *4a* | 0.5045(13) | 0.5045(13) | 0.5045(13) | 0.090 |
| $O_4$ | *4a* | 0.7366(8) | 0.7366(8) | 0.7366(8) | 0.090 |
| $O_5$ | *12b* | 0.2485(11) | 0.2610(10) | 0.0123(7) | 0.090 |
| $O_6$ | *12b* | 0.2554(12) | 0.2342(9) | 0.5424(6) | 0.090 |
| $O_7$ | *12b* | -0.0039(12) | -0.0071(10) | 0.2459(13) | 0.090 |
| $O_8$ | *12b* | -0.0044(12) | -0.0008(18) | 0.7594(10) | 0.090 |

**Table II.** Unit cell parameters of $Ce_2Zr_2O_8$ at different pressures and compressibilities (κ) in GPa$^{-1}$. $\kappa_x = \frac{-1}{x}\frac{\partial x}{\partial P}$

| Pressure (GPa) | *a* (Å) | V (Å$^3$) | $\kappa_a$ (GPa$^{-1}$) | $\kappa_V$ (GPa$^{-1}$) |
|---|---|---|---|---|
| Ambient | 10.5444(2) | 1172.34(4) | - | - |
| 3 | 10.4936(1) | 1155.51(2) | | |
| 4 | 10.4772(1) | 1150.09(2) | | |
| 5.1 | 10.4612(4) | 1144.83(8) | | |
| *0-5.1 GPa* | | | $1.54 \times 10^{-3}$ | $4.60 \times 10^{-3}$ |
| 7 | 10.4412(2) | 1138.27(4) | | |
| 8.4 | 10.4269(4) | 1133.62(8) | | |
| 10.3 | 10.4036(2) | 1126.04(4) | | |
| 12.1 | 10.3836(2) | 1119.56(4) | | |
| *7-12.1 GPa* | | | $1.08 \times 10^{-3}$ | $3.22 \times 10^{-3}$ |
| *0-12.1 GPa* | | | $1.26 \times 10^{-3}$ | $3.72 \times 10^{-3}$ |

**Table III:** Bulk modulus of related pyrochlores and fluorite-type materials.

| Compositions | $B_0$ (GPa) | $B_0$' | Reference |
|---|---|---|---|
| $Gd_2Zr_2O_7$ | 186 | - | [11] |
| $Gd_2Zr_2O_7$ | 153.4 | 10.5 | [11] |
| $La_2Hf_2O_7$ | 147.2 | 7.9 | [14] |
| $La_2Hf_2O_7$ | 179.3 | 4 | [14] |
| $CeO_2$ fluorite | 230 | - | [34] |
| Nano $CeO_2$ fluorite | 328 | - | [34] |
| $CeO_2$ (cotunite) | 304 | 4 | [35] |
| $ZrO_2$ (cotunite) | 332 | 2.3 | [35] |
| $Gd_2Zr_2O_7$ | 161.5 | 9 | [15] |
| $Gd_2Zr_2O_7$ | 156 | 7 | [20] |
| $Gd_2Zr_2O_7$ | 154 | 8 | [20] |
| $Nd_2Zr_2O_7$ | 145 | 11 | [20] |
| $Nd_2Zr_2O_7$ | 140 | 14 | [20] |
| $Ce_2Zr_2O_7$ | 252 | 0 | [20] |
| $Ce_2Zr_2O_7$ | 255 | 0 | [20] |
| $La_2Zr_2O_7$ | 200 | - | [48] |
| $La_2Hf_2O_7$ | 207 | - | [48] |
| $Y_2Zr_2O_7$ | 225 | - | [48] |
| $Y_2Hf_2O_7$ | 238 | - | [48] |

**Table IV:** Position coordinates of rhombohedral $Ce_2Zr_2O_8$ ($P3_2$, No 145) as obtained from transformed cubic cell. Unit-cell parameters are obtained by Le Bail refinements $a_h$ = 14.6791 Å , $c_h$ = 17.9421 Å, V = 3348.13 Å$^3$, Z = 24.

| Atoms | Wyckoff positions | x | y | z | Occupation |
|---|---|---|---|---|---|
| $Ce_1$ | *3a* | 0.333(3) | 0.333(3) | 0.129(2) | 1 |
| $Ce_2$ | *3a* | 0.082(3) | 0.331(3) | 0.626(2) | 1 |
| $Ce_3$ | *3a* | 0.338(3) | 0.086(3) | 0.626(2) | 1 |
| $Ce_4$ | *3a* | 0.581(3) | 0.585(3) | 0.626(2) | 1 |
| $Ce_5$ | *3a* | 0.171(3) | 0.251(3) | 0.290(2) | 1 |
| $Ce_6$ | *3a* | 0.416(3) | 0.254(3) | 0.291(2) | 1 |
| $Ce_7$ | *3a* | 0.414(3) | 0.497(3) | 0.290(2) | 1 |
| $Ce_8$ | *3a* | 0.082(3) | 0.087(3) | 0.626(2) | 1 |
| $Ce_9$ | *3a* | 0.581(3) | 0.329(3) | 0.626(2) | 1 |
| $Ce_{10}$ | *3a* | 0.339(3) | 0.586(3) | 0.626(2) | 1 |
| $Ce_{11}$ | *3a* | 0.583(3) | 0.830(3) | 0.624(2) | 1 |
| $Ce_{12}$ | *3a* | 0.838(3) | 0.088(3) | 0.624(2) | 1 |
| $Ce_{13}$ | *3a* | 0.579(3) | 0.085(3) | 0.624(2) | 1 |
| $Ce_{14}$ | *3a* | 0.499(3) | 0.167(3) | 0.464(2) | 1 |
| $Ce_{15}$ | *3a* | 0.500(3) | 0.665(3) | 0.464(2) | 1 |
| $Ce_{16}$ | *3a* | 0.002(3) | 0.168(3) | 0.464(2) | 1 |
| $Zr_1$ | *3a* | 0.334(2) | 0.334(2) | 0.6277(16) | 1 |
| $Zr_2$ | *3a* | 0.749(2) | 0.664(2) | 0.4580(17) | 1 |
| $Zr_3$ | *3a* | 0.003(2) | 0.418(2) | 0.4581(17) | 1 |
| $Zr_4$ | *3a* | 0.249(2) | 0.919(2) | 0.4582(17) | 1 |
| $Zr_5$ | *3a* | 0.170(2) | 0.253(2) | 0.7891(17) | 1 |
| $Zr_6$ | *3a* | 0.414(2) | 0.250(2) | 0.7888(17) | 1 |
| $Zr_7$ | *3a* | 0.417(2) | 0.498(2) | 0.7891(17) | 1 |
| $Zr_8$ | *3a* | 0.752(2) | 0.422(2) | 0.4580(17) | 1 |
| $Zr_9$ | *3a* | 0.245(2) | 0.663(2) | 0.4578(17) | 1 |
| $Zr_{10}$ | *3a* | 0.004(2) | 0.916(2) | 0.458(2) | 1 |
| $Zr_{11}$ | *3a* | 0.250(2) | 0.162(2) | 0.4607(17) | 1 |
| $Zr_{12}$ | *3a* | 0.506(2) | 0.421(2) | 0.4604(17) | 1 |
| $Zr_{13}$ | *3a* | 0.246(2) | 0.417(2) | 0.4605(17) | 1 |
| $Zr_{14}$ | *3a* | 0.163(2) | 0.496(2) | 0.2949(16) | 1 |
| $Zr_{15}$ | *3a* | 0.172(2) | 0.000(2) | 0.2946(17) | 1 |
| $Zr_{16}$ | *3a* | 0.666(2) | 0.504(2) | 0.2949(16) | 1 |
| $O_{1a}$ | *3a* | 0.335(5) | 0.332(5) | 0.000(3) | 1 |
| $O_{1b}$ | *3a* | 0.499(5) | 0.168(5) | 0.337(3) | 1 |
| $O_{1c}$ | *3a* | 0.497(5) | 0.666(5) | 0.337(3) | 1 |
| $O_{1d}$ | *3a* | 0.002(5) | 0.166(5) | 0.336(3) | 1 |
| $O_{2a}$ | *3a* | 0.333(5) | 0.335(5) | 0.256(3) | 1 |
| $O_{2b}$ | *3a* | 0.998(5) | 0.162(5) | 0.584(3) | 1 |
| $O_{2c}$ | *3a* | 0.506(5) | 0.171(5) | 0.584(3) | 1 |
| $O_{2d}$ | *3a* | 0.498(5) | 0.670(5) | 0.585(3) | 1 |

| Atoms | Wyckoff positions | x | y | z | Occupation |
|---|---|---|---|---|---|
| $O_{3a}$ | *3a* | 0.334(3) | 0.334(3) | 0.505(2) | 1 |
| $O_{3b}$ | *3a* | 0.831(3) | 0.828(3) | 0.498(2) | 1 |
| $O_{3c}$ | *3a* | 0.840(3) | 0.336(3) | 0.499(2) | 1 |
| $O_{3d}$ | *3a* | 0.330(3) | 0.836(3) | 0.499(2) | 1 |
| $O_{4a}$ | *3a* | 0.334(4) | 0.334(4) | 0.736(2) | 1 |
| $O_{4b}$ | *3a* | 0.677(3) | 0.519(3) | 0.422(2) | 1 |
| $O_{4c}$ | *3a* | 0.148(3) | 0.491(3) | 0.422(2) | 1 |
| $O_{4d}$ | *3a* | 0.177(3) | 0.991(3) | 0.422(2) | 1 |
| $O_{5a}$ | *3a* | 0.084(4) | 0.846(4) | 0.499(3) | 1 |
| $O_{5b}$ | *3a* | 0.821(4) | 0.572(4) | 0.499(3) | 1 |
| $O_{5c}$ | *3a* | 0.095(4) | 0.583(4) | 0.498(3) | 1 |
| $O_{5d}$ | *3a* | 0.092(4) | 0.341(4) | 0.494(3) | 1 |
| $O_{5e}$ | *3a* | 0.326(4) | 0.085(4) | 0.494(3) | 1 |
| $O_{5f}$ | *3a* | 0.583(4) | 0.574(4) | 0.494(3) | 1 |
| $O_{5g}$ | *3a* | 0.409(5) | 0.494(4) | 0.676(2) | 1 |
| $O_{5h}$ | *3a* | 0.173(4) | 0.250(5) | 0.676(2) | 1 |
| $O_{5i}$ | *3a* | 0.418(5) | 0.258(5) | 0.675(2) | 1 |
| $O_{5j}$ | *3a* | 0.747(5) | 0.654(4) | 0.335(3) | 1 |
| $O_{5k}$ | *3a* | 0.014(4) | 0.428(5) | 0.335(3) | 1 |
| $O_{5l}$ | *3a* | 0.240(5) | 0.919(5) | 0.335(3) | 1 |
| $O_{6a}$ | *3a* | 0.245(4) | 0.135(3) | 0.341(2) | 1 |
| $O_{6b}$ | *3a* | 0.531(4) | 0.444(5) | 0.341(3) | 1 |
| $O_{6c}$ | *3a* | 0.224(5) | 0.422(5) | 0.341(3) | 1 |
| $O_{6d}$ | *3a* | 0.229(5) | 0.646(4) | 0.352(2) | 1 |
| $O_{6e}$ | *3a* | 0.021(4) | 0.917(5) | 0.351(5) | 1 |
| $O_{6f}$ | *3a* | 0.750(5) | 0.438(5) | 0.352(2) | 1 |
| $O_{6g}$ | *3a* | 0.268(3) | 0.185(4) | 0.814(4) | 1 |
| $O_{6h}$ | *3a* | 0.481(4) | 0.416(4) | 0.814(4) | 1 |
| $O_{6i}$ | *3a* | 0.251(4) | 0.399(3) | 0.814(4) | 1 |
| $O_{6j}$ | *3a* | 0.592(4) | 0.366(5) | 0.495(3) | 1 |
| $O_{6k}$ | *3a* | 0.301(5) | 0.559(4) | 0.495(3) | 1 |
| $O_{6l}$ | *3a* | 0.109(4) | 0.076(4) | 0.495(3) | 1 |
| $O_{7a}$ | *3a* | 0.579(4) | 0.829(4) | 0.748(3) | 1 |
| $O_{7b}$ | *3a* | 0.838(4) | 0.083(4) | 0.747(3) | 1 |
| $O_{7c}$ | *3a* | 0.583(4) | 0.087(4) | 0.747(3) | 1 |
| $O_{7d}$ | *3a* | 0.421(4) | 0.004(4) | 0.423(3) | 1 |
| $O_{7e}$ | *3a* | 0.664(4) | 0.751(4) | 0.423(3) | 1 |
| $O_{7f}$ | *3a* | 0.916(4) | 0.246(4) | 0.423(3) | 1 |
| $O_{7g}$ | *3a* | 0.090(5) | 0.336(4) | 0.252(3) | 1 |

| **Atoms** | **Wyckoff positions** | **x** | **y** | **z** | **Occupation** |
|---|---|---|---|---|---|
| $O_{7h}$ | *3a* | 0.332(4) | 0.088(4) | 0.251(3) | 1 |
| $O_{7i}$ | *3a* | 0.579(4) | 0.577(5) | 0.251(3) | 1 |
| $O_{7j}$ | *3a* | 0.245(4) | 0.166(4) | 0.582(3) | 1 |
| $O_{7k}$ | *3a* | 0.501(4) | 0.411(5) | 0.582(3) | 1 |
| $O_{7l}$ | *3a* | 0.257(5) | 0.424(4) | 0.582(3) | 1 |
| $O_{8a}$ | *3a* | 0.739(4) | 0.154(4) | 0.588(3) | 1 |
| $O_{8b}$ | *3a* | 0.513(4) | 0.919(4) | 0.588(3) | 1 |
| $O_{8c}$ | *3a* | 0.749(4) | 0.929(4) | 0.588(3) | 1 |
| $O_{8d}$ | *3a* | 0.583(4) | 0.324(4) | 0.258(3) | 1 |
| $O_{8e}$ | *3a* | 0.343(4) | 0.594(4) | 0.257(3) | 1 |
| $O_{8f}$ | *3a* | 0.073(4) | 0.082(4) | 0.258(3) | 1 |
| $O_{8g}$ | *3a* | 0.927(4) | 0.015(4) | 0.416(3) | 1 |
| $O_{8h}$ | *3a* | 0.652(4) | 0.245(4) | 0.415(3) | 1 |
| $O_{8i}$ | *3a* | 0.422(4) | 0.740(4) | 0.416(3) | 1 |
| $O_{8j}$ | *3a* | 0.419(4) | 0.509(4) | 0.410(3) | 1 |
| $O_{8k}$ | *3a* | 0.158(4) | 0.244(4) | 0.409(3) | 1 |
| $O_{8l}$ | *3a* | 0.424(4) | 0.248(4) | 0.409(3) | 1 |

**Table V:** Ambient-pressure Raman frequencies (ω), pressure coefficients (dω/dP), and mode Grüneisen parameters (γ) for the cubic phase of $Ce_2Zr_2O_8$.

| ω ($cm^{-1}$) | dω/dP ($cm^{-1}$/GPa) | γ | ω ($cm^{-1}$) | dω/dP ($cm^{-1}$/GPa) | γ |
|---|---|---|---|---|---|
| 68 | 0.10 | 0.31 | 363 | 1.38 | 0.81 |
| 89 | 0.45 | 1.08 | 398 | 3.74 | 2.01 |
| 121 | 0.78 | 1.38 | 440 | 3.78 | 1.84 |
| 142 | 0.34 | 0.51 | 471 | 3.66 | 1.66 |
| 171 | 1.25 | 1.56 | 520 | 3.24 | 1.33 |
| 197 | 4.70 | 5.11 | 549 | 4.86 | 1.89 |
| 231 | 1.25 | 1.16 | 560 | 6.22 | 2.38 |
| 259 | 0.70 | 0.58 | 593 | 5.08 | 1.83 |
| 277 | 2.02 | 1.56 | 643 | 6.50 | 2.16 |
| 292 | 2.94 | 2.15 | 759 | 5.30 | 1.49 |
| 308 | 3.54 | 2.46 | | | |

**Table VI:** Raman frequencies and pressure coefficients for the rhombohedral phase of $Ce_2Zr_2O_8$ determined at 8.7 GPa.

| ω ($cm^{-1}$) | dω/dP ($cm^{-1}$/GPa) | ω ($cm^{-1}$) | dω/dP ($cm^{-1}$/GPa) | ω ($cm^{-1}$) | dω/dP ($cm^{-1}$/GPa) |
|---|---|---|---|---|---|
| 68.8 | 0.10 | 303.8 | 2.33 | 565.4 | 3.89 |
| 85.4 | 0.76 | 315.2 | 2.27 | 595.2 | 3.92 |
| 1200.7 | -1.45 | 328.4 | 2.85 | 616 | 2.32 |
| 114.7 | 0.95 | 365.8 | 0.57 | 629.3 | 1.38 |
| 125.2 | 0.71 | 387.5 | 0.75 | 669.6 | 0.74 |
| 167.4 | 0.79 | 402.7 | 3.16 | 687.5 | 0.73 |
| 186.1 | 1.01 | 425.2 | 2.39 | 710.3 | 0.69 |
| 206 | 1.21 | 449.5 | 2.44 | 733.7 | 1.61 |
| 232.4 | 2.64 | 466.5 | 2.62 | 778.8 | 2.10 |
| 250.2 | 2.18 | 497.5 | 3.18 | 850.2 | 3.82 |
| 276.6 | 2.62 | 518.3 | 3.61 | | |
| 291.5 | 1.92 | 546.4 | 3.17 | | |

## Figure Captions

**Fig. 1.** (color online) Schematic view of the room-temperature low-pressure (cubic) and high-pressure (rhombohdral) phases of $Ce_2Zr_2O_8$.

**Fig. 2:** XRD patterns of $Ce_2Zr_2O_8$ at selected pressures. At 3 GPa the most intense peaks of the low-pressure phase are labeled. Most intense peaks characteristic of the high-pressure phase are indicated with dotted lines.

**Fig. 3.** (color online) XRD pattern of $Ce_2Zr_2O_8$ at 3 GPa, refined as cubic, $P2_13$, $a$ = 10.4936(1) Å. Experiments (circles) are shown together with the calculated patterns and residuals. Ticks indicate the position of Bragg reflections.

**Fig. 4.** Pressure-volume data of $Ce_2Zr_2O_8$ (circles). The solid line represents the calculated EOS.

**Fig. 5.** Raman spectra of $Ce_2Zr_2O_8$ at selected pressures. At ambient pressure the phonons assigned are indicated. Changes in the spectra indicate the occurrence of a phase transition at 5.5 GPa. The upper trace shows the spectrum collected at ambient pressure after pressure release.

**Fig. 6.** (color online) XRD pattern of $Ce_2Zr_2O_8$ at 12.1 GPa, refined as cubic, $P2_13$, $a$ = 10.3836(2) Å. Differences at the new weak peaks are shown as inset. Peak at 11.8 can be assigned to neon (111) reflection.

**Fig. 7.** (color online) XRD pattern of $Ce_2Zr_2O_8$ at 12.1 GPa, refined as rhombohedral, $P3_2$, $a$ = 14.6791(3) Å, $c$ = 17.9421(5) Å. Neon (111) overlaps with the peaks from the rhombohedral $Ce_2Zr_2O_8$.

**Fig. 8.** Raman modes of low- and high-pressure phases of $Ce_2Ze_2O_8$ as a function of pressure. Solid symbols: cubic phase. Empty symbols: rhombohedral phase. Solid lines are linear fits in the low-P phase. B-spline curves are used for the HP phase.

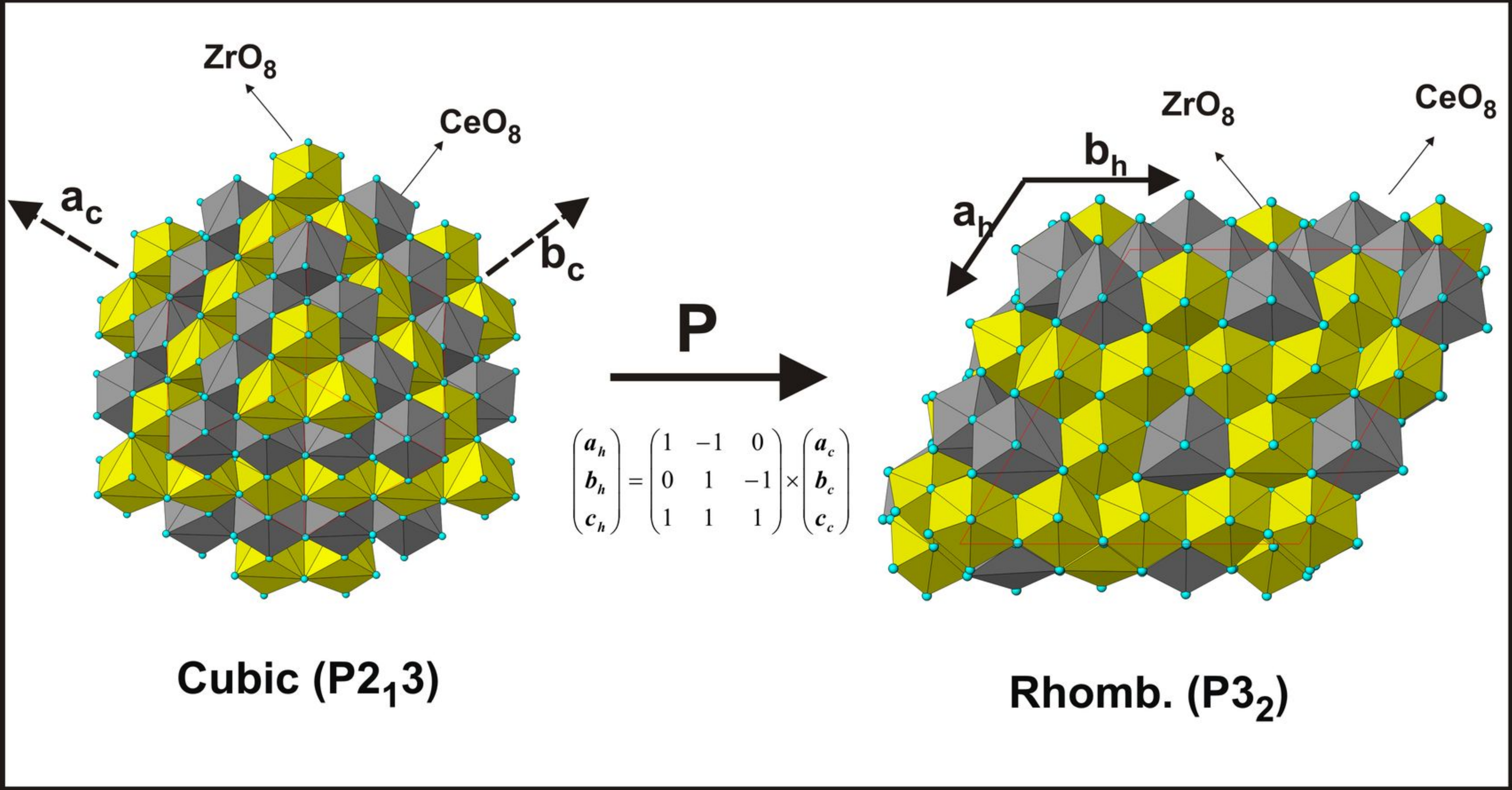
ZrO8
CeO8
a_c
b_c
P
a_h
b_h
ZrO8
CeO8
(a_h, b_h, c_h) = ((1, −1, 0), (0, 1, −1), (1, 1, 1)) × (a_c, b_c, c_c)
Cubic (P2_1 3)
Rhomb. (P3_2)

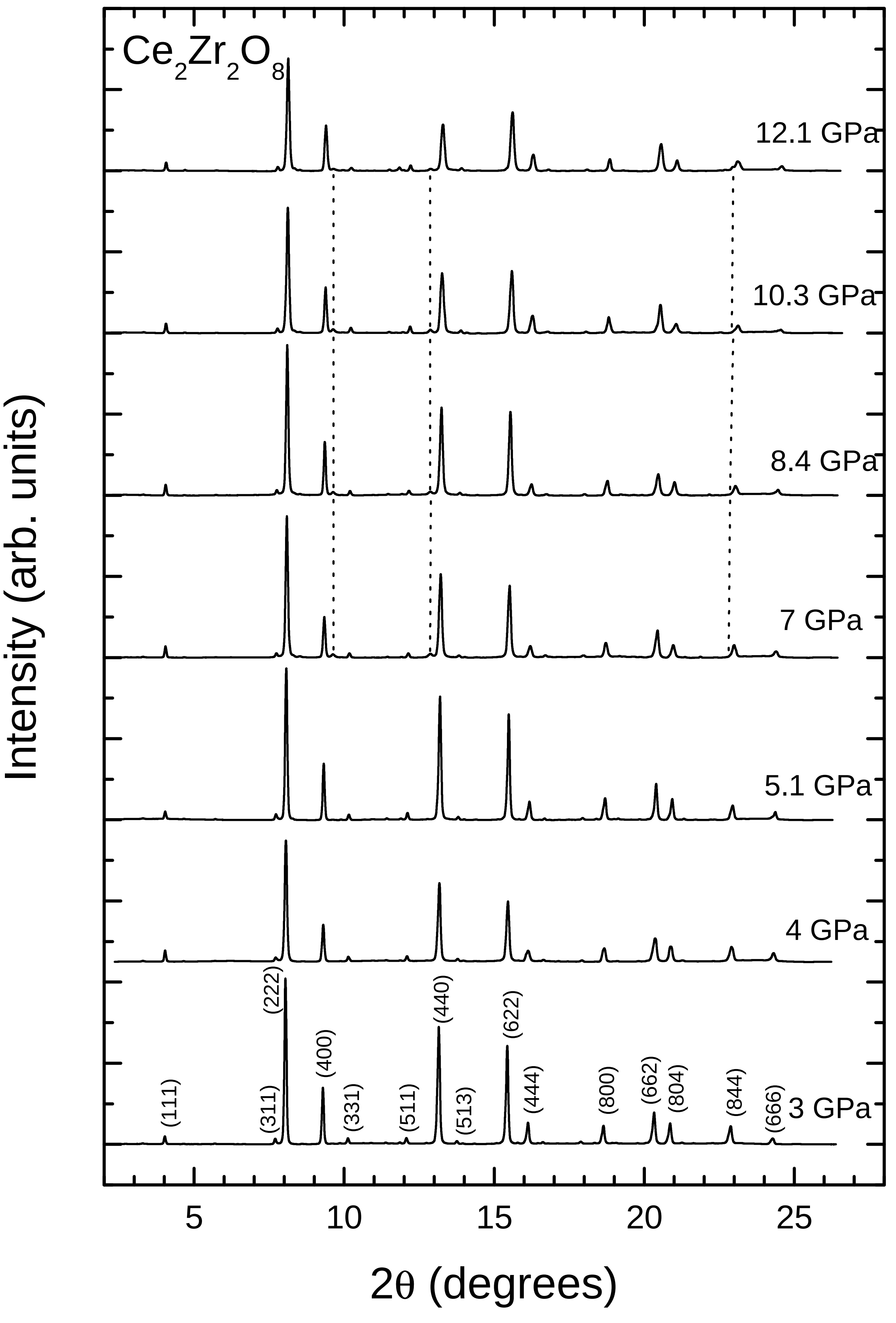

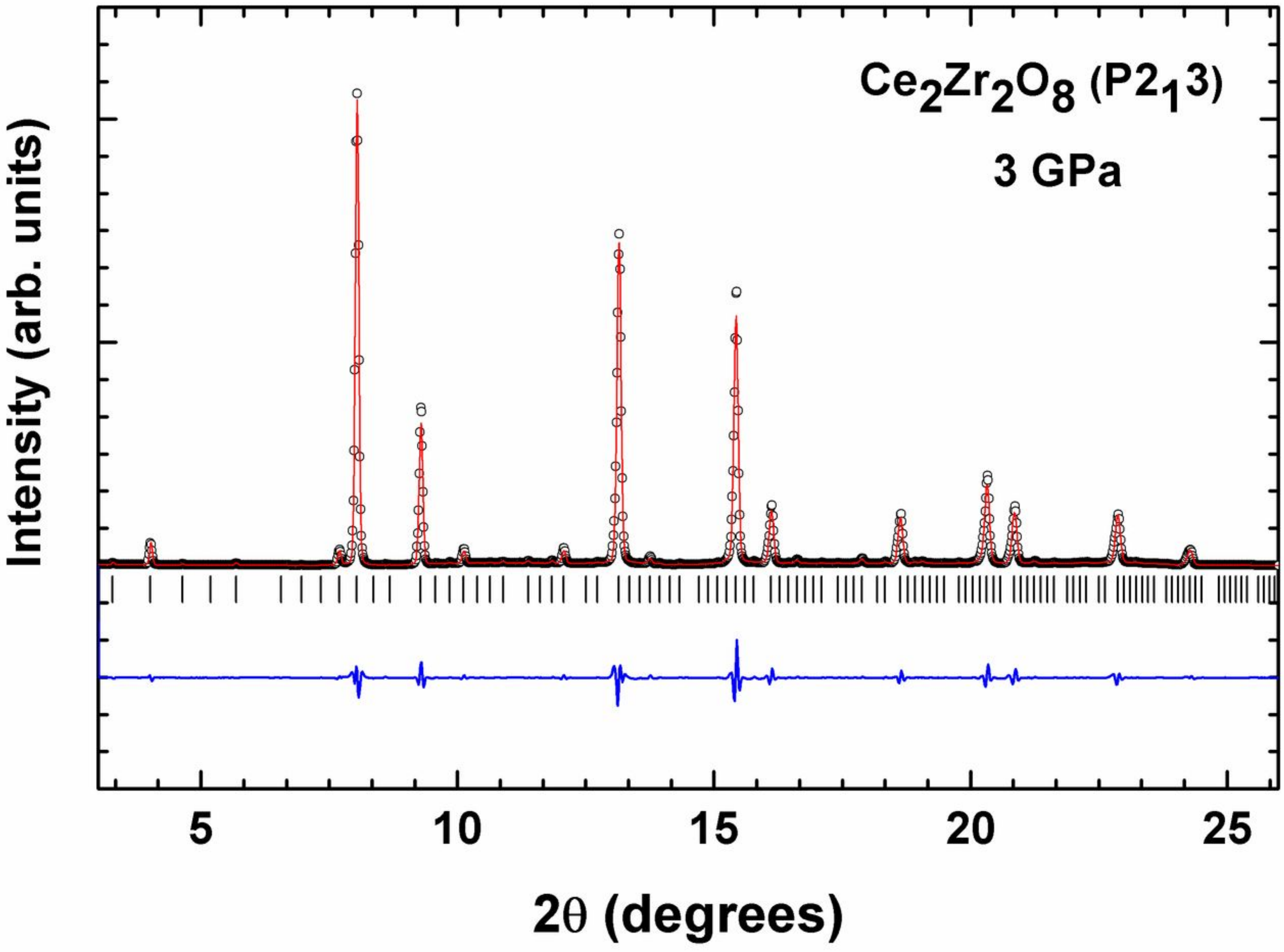

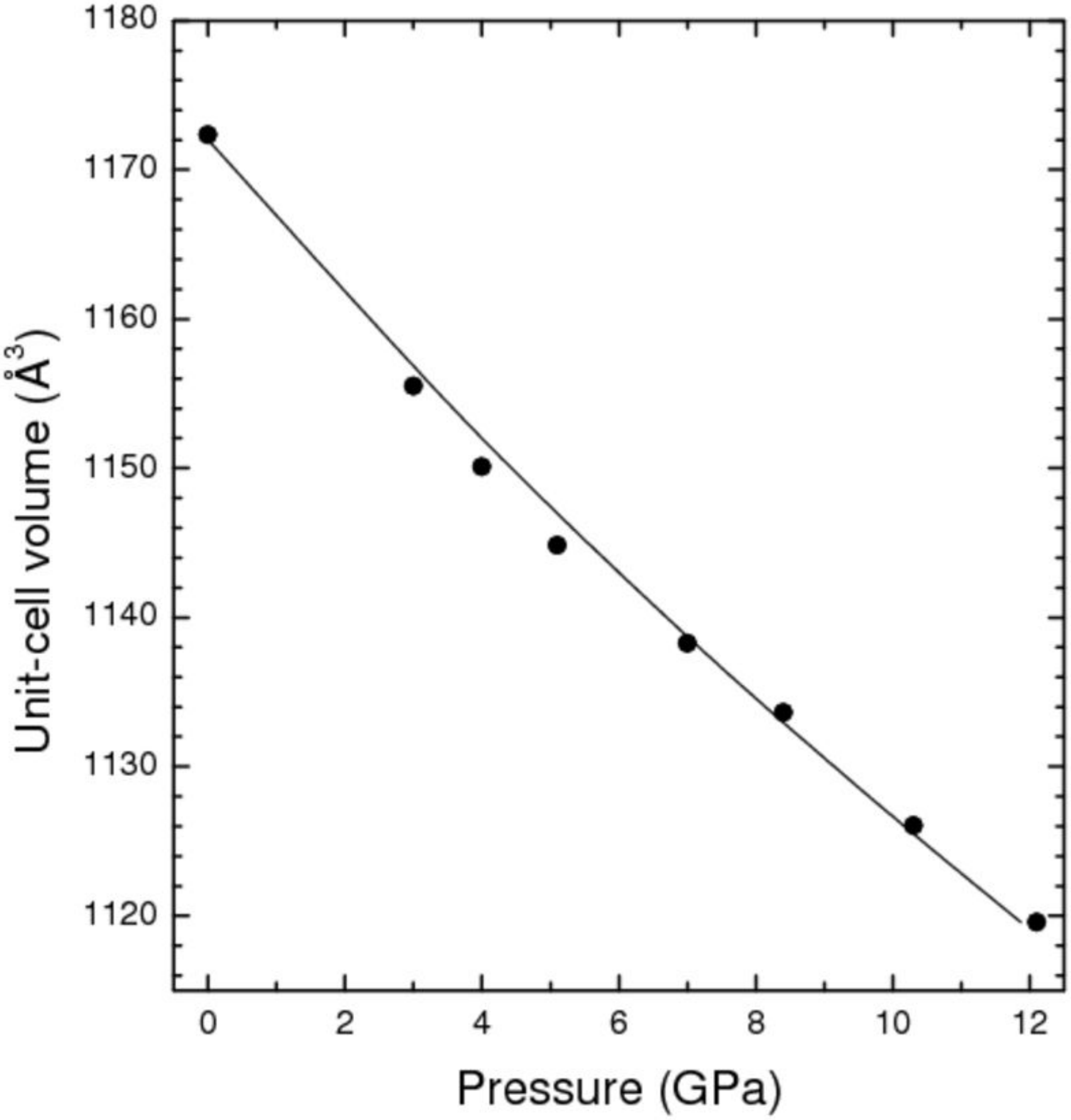

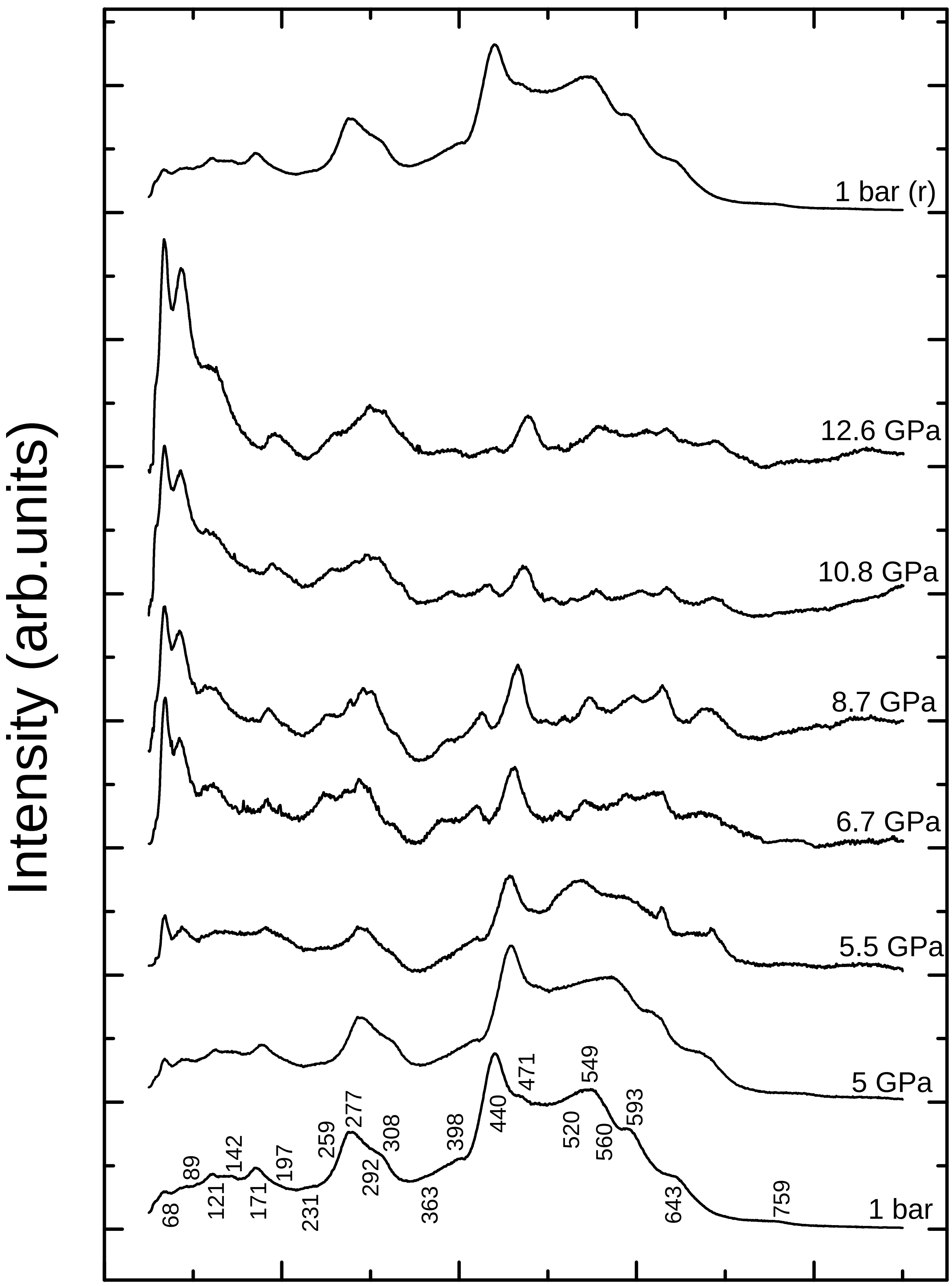

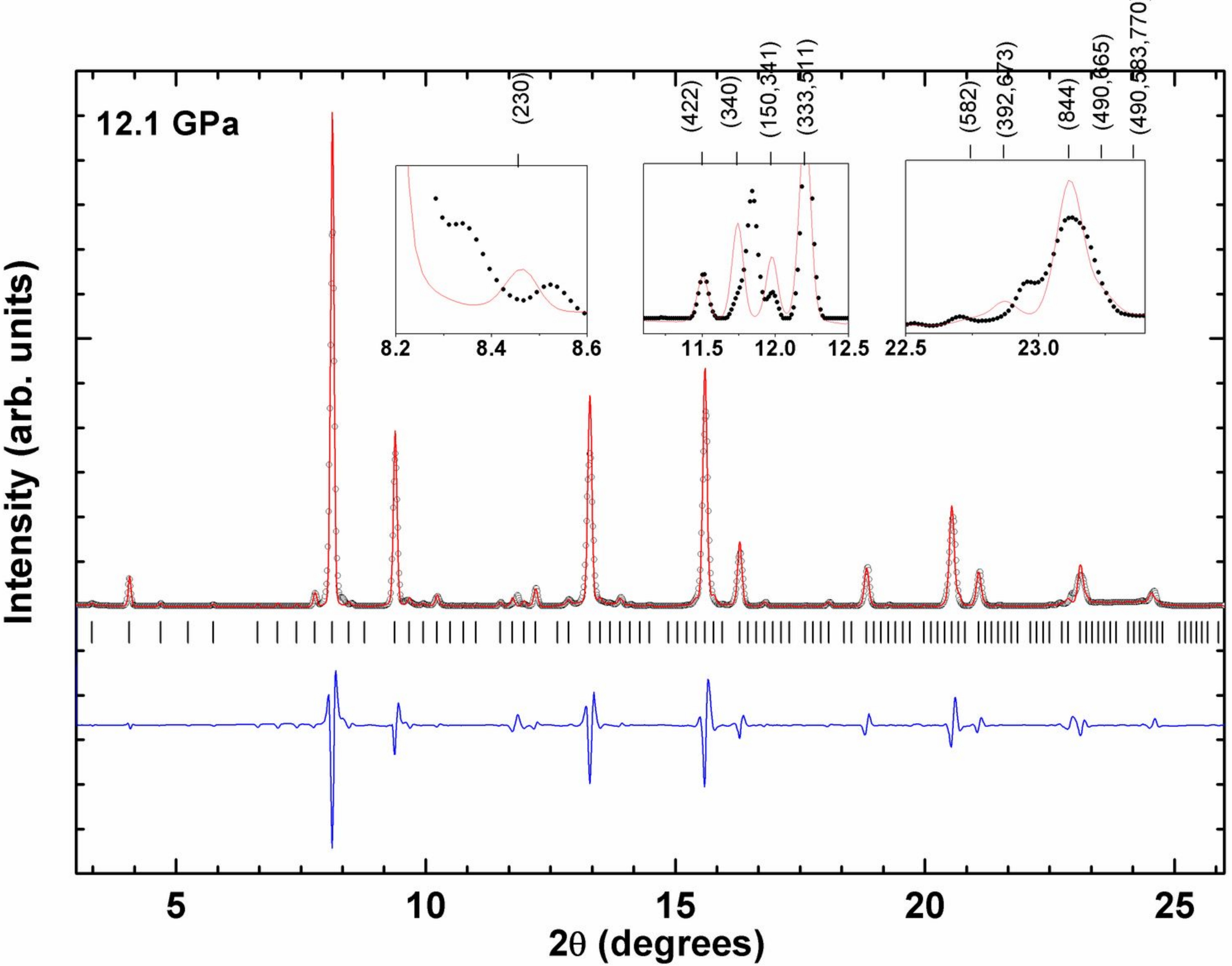

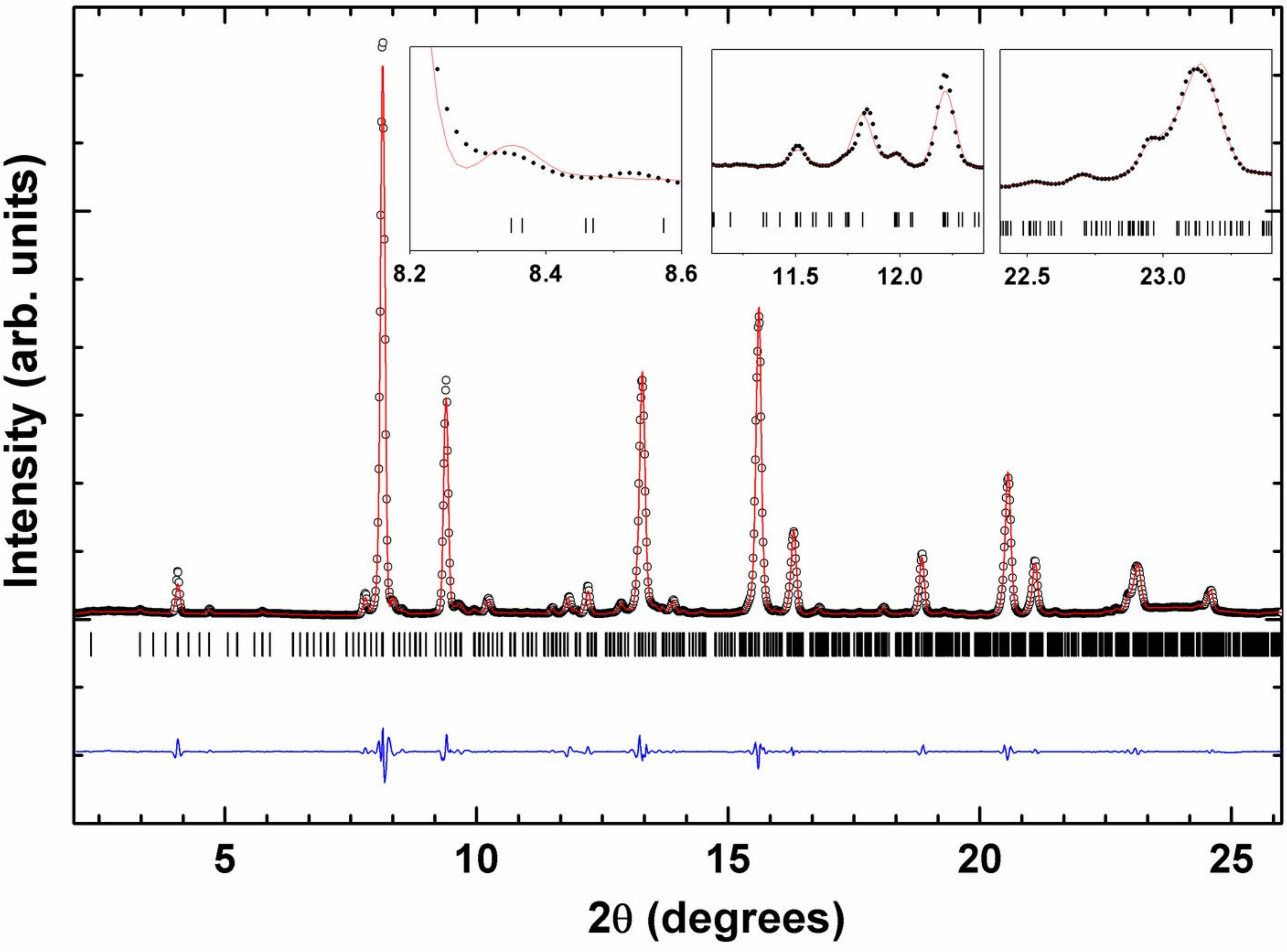

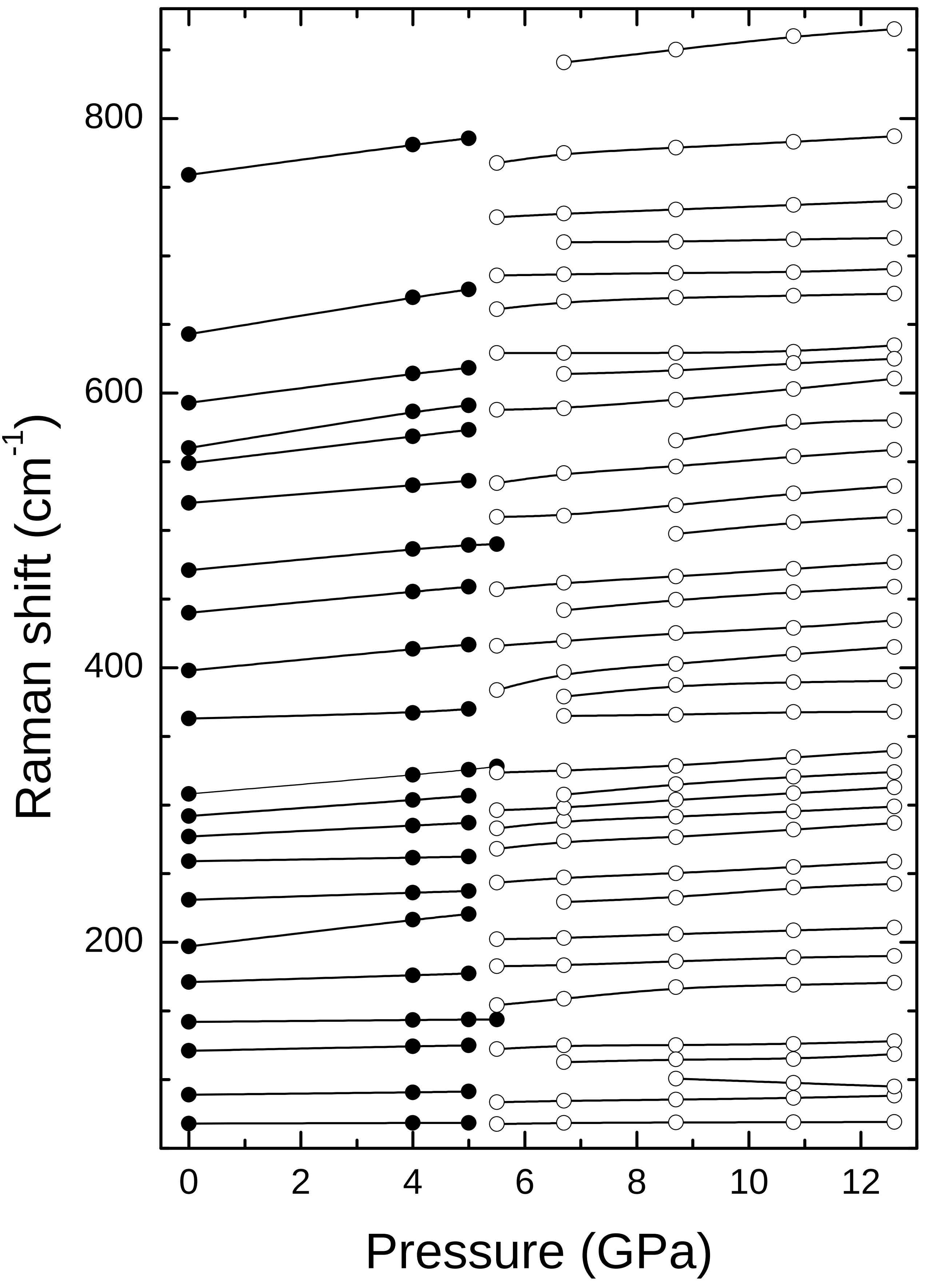